\documentstyle[preprint,eqsecnum,aps,epsf]{revtex}

\begin{document}


\newcommand{\bm}[1]{
\mbox{\boldmath$ #1 $}}
\def\M{{\cal M}}
\def\G{{\cal G}}
\def\H{{\cal H}}
\def\N{{\cal N}}
\def\V{{\cal V}}
\def\D{{\cal D}}
\def\R{{\cal R}}
\def\Z{{\cal Z}}
\def\v{{\bm v}}
\def\x{{\bm x}}

\newtheorem{theorem}{Theorem}
\newtheorem{lemma}[theorem]{Lemma}
\newtheorem{prop}{Proposition}
\newtheorem{definition}[theorem]{Definition}
\newenvironment{proof}{\par\noindent {\bf Proof}\par}{\par\medskip}
\def\QED{$\Box$}

\preprint{YITP-97-23}
\title{
State after quantum tunneling with gravity
}
\author{
Shinji Mukohyama
}
\address{
Yukawa Institute for Theoretical Physics, Kyoto University \\ 
Kyoto 606-01, Japan \\
}
\date{\today}

\maketitle


\begin{abstract} 

The Wheeler-DeWitt equation is investigated and used to examine a
state after a quantum tunneling with gravity. To make arguments
definite we treat a discretized version of the Wheeler-DeWitt equation
and adopt the WKB method. We expand an Euclidean wave function around
an instanton, by  using a deviation equation of a vector field tangent
to a congruence of instantons. The instanton around which we expand
the wave function corresponds to a so-called most probable escape path
(MPEP). It is shown that, when the wave function is analytically
continued, the corresponding state of physical perturbations is
equivalent to the vacuum state determined by positive-frequency mode
functions which satisfy appropriate boundary conditions. Thus a
quantum field theory is effective to investigate a state after a
quantum tunneling with gravity. The effective Lagrangian describing
the field theory is obtained by simply reducing the original
Lagrangian to a subspace spanned by the physical perturbations. The
result of this paper does not depend on the operator ordering and can
be applied to all physical perturbations, including gravitational
perturbations, around a general MPEP. 

\end{abstract}

\vfill
\newpage



\section{Introduction}
\label{sec:Introduction}

The universe is thought to have been experienced phase transitions
many times. Some of these phase transitions may be due to quantum
tunneling. Hence there arises an intellectual and practical urge to
know a state after a quantum tunneling. For this purpose many authors
investigated the Schrodinger equation of scalar fields in flat
spacetime~\cite{Gervais&Sakita,Tunneling} in the multi-dimensional
tunneling approach~\cite{MPEP}. They showed that a state after a
quantum tunneling in flat spacetime is the vacuum state determined by 
positive-frequency mode functions which satisfy appropriate boundary
conditions. The result says that a quantum field theory of the scalar
fields is effective to seek a particle spectrum after a quantum
tunneling if the boundary conditions are attached on the mode
functions. The effectiveness makes it possible to calculate a CMB
anisotropy of the universe theoretically in some inflationary
scenarios~\cite{CMB-calc}.

However the calculated anisotropy is based on the formalism developed in
flat spacetime, as mentioned above. Since gravitational effects can be
significant in some cases~\cite{Colemen&Luccia}, we want to generalize
the formalism to include gravity. As far as only the scalar fields'
perturbations are concerned, the generalization was
done in Ref.~\cite{Grav-effects}, by neglecting gravitational
perturbations. In this case, a state of the scalar fields'
perturbations after a quantum tunneling with gravity is obtained by
the same method as in flat spacetime, except evolution equations of
mode functions is replaced by those on the curved background. Thus a
quantum field theory of the scalar fields in curved spacetime is
effective to investigate a state of the scalar fields' perturbations
after a quantum tunneling, provided that gravitational perturbations
are neglected. However, when we intend to investigate a state of
gravitational perturbations after a quantum tunneling (for example,
when we calculate a spectrum of the primordial gravitational waves),
we have to include gravitational perturbations. The corresponding 
formalism must be derived from a more fundamental level.

When we investigate a classical dynamics of gravitational
perturbations, so-called gauge invariant variables play a important
role. The gauge invariant variables are those linear combinations of
perturbations which are invariant to linear order under coordinate
transformations, and a maximal set of them describes all physical
degrees of freedom of linear perturbations. For several background
geometries, it was shown that classical dynamics of the gauge 
invariant variables can be described by some auxiliary scalar fields
(Ref.~\cite{Gravitation} for Minkowski background,
Ref.~\cite{Ford&Parker} for the Robertson-Walker background,
Ref.~\cite{Milne} for the Milne background, and Ref.~\cite{Grav-wave}
for the one-bubble inflationary background). Hence we can expect that
a quantum field theory of the auxiliary scalar fields can be applied
to investigate a state of gravitational perturbations after a quantum
tunneling, although the equivalence is established only in a
classical level.

At this point, we can expect that a quantum field theory will be
effective to investigate a state of all physical perturbations after a
quantum tunneling with gravity. Thus we want to show from a quantum
theory of gravity that the expectation is true. In all papers with
this aim~\cite{Banks-etc}, the Wheeler-DeWitt equation is expanded by
some collective coordinates around a background. Our strategy in this
paper is to expand not the Wheeler-DeWitt equation but a wave function 
itself covariantly. Anyway, due to the result of this paper, we can
safely use the quantum field theory of perturbations in order to
calculate a CMB anisotropy, a spectrum of the primordial
gravitational waves, etc. in some inflationary scenarios.

Now let us review a quantum theory of gravity based on a canonical
formalism. Consider a system of a $n$-dimensional spacetime
$(M,g_{\mu\nu})$ without boundary and scalar fields $\chi^a$ described
by the action 
\begin{equation}
 I_{E-KG} = \int d^nx\sqrt{-g}\left[ \frac{1}{2\kappa^2}R - 
	\frac{1}{2}g^{\mu\nu}\gamma_{ab}(\chi^c)
	\partial_{\mu}\chi^a\partial_{\nu}\chi^b - V(\chi^a)\right]
	\ \ , 
\end{equation}
where $\gamma_{ab}$ is an arbitrary positive-definite matrix. In the
ADM decomposition~\cite{Gravitation} of the spacetime  
\begin{equation}
 ds^2 = -\Lambda_{\perp}^2dt^2 + 
	q_{ij}(dx^i + \Lambda^idt)(dx^j + \Lambda^jdt)\ \ ,
\end{equation}
functional derivatives of the action $I_{E-KG}$ with respect to
$\Lambda_{\perp}$ and $\Lambda^i$ give the following constraints. 
\begin{eqnarray}
 \H_{\perp\x} & = & 0 \ \ ,\nonumber\\
 \H_{i\x} & = & 0 \ \ ,
\end{eqnarray}
where 
\begin{eqnarray}
 \H_{\perp\x} & \equiv & \kappa^2G_{ijkl}\pi^{ij}\pi^{kl}
	+ \frac{1}{2\sqrt{q}}\gamma^{ab}\pi_a\pi_b \nonumber\\
 & &	\left.
	+ \sqrt{q}\left[ -\frac{{}^{(n-1)}R}{2\kappa^2}
	+\frac{1}{2}q^{ij}\gamma_{ab}(\chi^c)
	\partial_i\chi^a\partial_j\chi^b 
	+ V(\chi^a)\right]\right|_{\x}
	\ \ ,\nonumber\\
 \H_{i\x} & \equiv & \left.
	- 2\sqrt{q}D_j\left(\frac{\pi^j_i}{\sqrt{q}}\right) +
	\pi_a\partial_i\chi^a\right|_{\x}\ \ .
\end{eqnarray}
The former is called Hamiltonian constraints and the later is called
momentum constraints. In the expressions, $\pi^{ij}$ is a momentum
conjugate to $q_{ij}$, $\pi_a$ is one conjugate to $\chi^a$, $D_j$
denotes a covariant derivative compatible with $q_{ij}$, and  
\begin{equation}
 G_{ijkl} \equiv \frac{1}{\sqrt{q}}\left(
	q_{ik}q_{jl} + q_{il}q_{jk} -\frac{2}{n-2}q_{ij}q_{kl}\right) 
\end{equation}
for $n\ne 2$, 
\begin{equation}
 G_{ijkl} \equiv \frac{1}{\sqrt{q}}\left(
	q_{ik}q_{jl} + q_{il}q_{jk}\right) \label{eqn:super-metric}
\end{equation}
for $n=2$. It is well-known that Poisson brackets among the above
constraints are given by 
\begin{eqnarray}
 \{f\H_{\perp},g\H_{\perp}\}_P & = & (fD^ig-gD^if)\H_i
		\ \ ,\nonumber\\
 \{f^i\H_{i},g\H_{\perp}\}_P & = & (f^i\partial_ig)\H_{\perp}
		\ \ ,\nonumber\\
 \{f^i\H_{i},g^j\H_{j}\}_P & = & (f^jD_jg^i-g^jD_jf^i)\H_i\ \ ,
	\label{eqn:class-Dirac-algebra}
\end{eqnarray}
where $f\H_{\perp}$ and $f^i\H_i$ denote $\int d\x f(\x)\H_{\perp\x}$
and $\int d\x f^i(\x)\H_{i\x}$, respectively. These show that  
the constraints are first class. A method to quantize a system with
first class constraints was given by Dirac~\cite{Dirac} and leads the
following simultaneous differential equations for a wave functional
$\Psi$. 
\begin{eqnarray}
 \hat{\H}_{\perp\x}\Psi & = & 0 \ \ ,\nonumber\\	
 \hat{\H}_{i\x}\Psi & = & 0 \ \ ,\label{eqn:WD-eq}
\end{eqnarray}
where $\hat{\H}_{\perp\x}$ and $\hat{\H}_{i\x}$ are differential
operators obtained by replacing $\pi^{ij}$ with $-i\hbar\delta/\delta
q_{ij}$, $\pi_a$ with $-i\hbar\delta/\delta\chi^a$ in $\H_{\perp\x}$
and $\H_{i\x}$, respectively. Equation (\ref{eqn:WD-eq}) is called the 
Wheeler-DeWitt equation~\cite{DeWitt}. In defining the differential
operators, there is a problem of operator ordering. The first
principle to determine the operator ordering is requiring the
following algebra~\cite{Ordering-problem}:
\begin{eqnarray}
 \left[f\hat{\H}_{\perp},g\hat{\H}_{\perp}\right]\Psi & = & 
	i\hbar (fD^ig-gD^if)\hat{\H}_i\Psi 	\ \ ,
				\label{eqn:Dirac-algebra1}\\
 \left[f^i\hat{\H}_{i},g\hat{\H}_{\perp}\right]\Psi & = & 
	i\hbar (f^i\partial_ig)\hat{\H}_{\perp}\Psi 	\ \ ,
				\label{eqn:Dirac-algebra2}\\
 \left[f^i\hat{\H}_{i},g^j\hat{\H}_{j}\right]\Psi & = & 
	i\hbar [f,g]^i\hat{\H}_i\Psi  \label{eqn:Dirac-algebra3} 
\end{eqnarray}
for an arbitrary functional $\Psi$. The algebra is a quantum version
of (\ref{eqn:class-Dirac-algebra}), and is called the Dirac
algebra~\cite{Dirac}. The condition
(\ref{eqn:Dirac-algebra1}-\ref{eqn:Dirac-algebra3}) is 
necessary in order for the equation (\ref{eqn:WD-eq}) to give a
consistent quantum theory.

The remaining part of this paper is organized as follows. In
Sec.~\ref{sec:WD-eq} we investigate a simultaneous differential
equations, which become the Wheeler-DeWitt equation in a limit. We
expand an Euclidean wave function around an instanton, by using a
deviation equation of a vector field tangent to a congruence of
instantons. The instanton around which we expand the wave function
corresponds to a so-called most probable escape path (MPEP). In
Sec.~\ref{sec:QFT}, by analytically continuing the wave function, it
is shown that the corresponding state of physical perturbations after
a quantum tunneling is the vacuum state determined by
positive-frequency mode functions which satisfy appropriate boundary
conditions. Hence a quantum field theory is effective to investigate a
state after a quantum tunneling. Sec.~\ref{sec:Sammary} is devoted to
summarize this paper.


\section{Euclidean wave function for a discretized Wheeler-DeWitt
equation}	\label{sec:WD-eq}

Let $(\M_x,G_{x\alpha\beta})$ be a family of $D$-dimensional
pseudo-Riemannian manifolds parameterized by an integer $x$
$(=1,2,\cdots,x_{max})$. A configuration space we consider is a
pseudo-Riemannian manifold constructed from them as 
\begin{equation}
 (\M,G_{\alpha\beta})=\oplus_x (\M_x,G_{x\alpha\beta})\ \ . 
			\label{eqn:direct-sum}
\end{equation}
Then we consider the following simultaneous differential equations for 
a complex function  $\Psi$ on $\M$. 
\begin{eqnarray}
 \hat{\H}_{\perp x} \Psi & = & 0 \ \ ,	\nonumber \\
 \hat{\H}_I \Psi & = & 0 \ \ , \label{eqn:constraint-eq}
\end{eqnarray}
where $x=1,2,\cdots,x_{max}$, $I=1,2,\cdots,I_{max}$ and 
\begin{eqnarray}
 \hat{\H}_{\perp x} & \equiv & -\frac{\hbar^2}{2}G_x^{\alpha\beta}
	D_{\alpha}D_{\beta} + V_x \ \ , \nonumber \\
 \hat{\H}_I & \equiv & -i\hbar v_I^{\alpha}D_{\alpha}
	\label{eqn:operators}
\end{eqnarray}
are linear differential operators on $\M$. As usual we call a solution
of this equation a {\it wave function}. In the expression $D$ is a
covariant derivative compatible with the metric $G_{\alpha\beta}$, 
$\{ V_x\}$ is a set of functions on $\M$, and $\{ v_I^{\alpha}\}$ is a
set of linearly independent vectors on $\M$. The Wheeler-DeWitt
equation (\ref{eqn:WD-eq}) can be written in this form provided that
it is properly discretized and the operator ordering problem is
ignored. $\hat{\H}_{\perp x}$ corresponds to the Hamiltonian
constraint at the point $x$ on a spacelike hypersurface, and
$\hat{\H}_I$ corresponds to the momentum constraint for a point and a
direction both of which are specified by $I$. For example, 
\begin{eqnarray}
 G_x^{q_{ij}(x')q_{kl}(x'')} & = & 
	2\kappa^2G_{ijkl}(x)\delta_{xx'}\delta_{xx''}\ \ ,\nonumber\\
 G_x^{q_{ij}(x')\chi^a(x'')} & = & 0 \ \ ,\nonumber\\
 G_x^{\chi^a(x')\chi^b(x'')} & = & 
	\frac{1}{\sqrt{q(x)}}\gamma^{ab}(x)\delta_{xx'}\delta_{xx''}
\end{eqnarray}
for the system considered in Sec. \ref{sec:Introduction}, where
$G_{ijkl}$ is defined by (\ref{eqn:super-metric}). We mention that
terms linear in the derivative, which may appear when the operator
ordering problem is solved, were not included in $\H_{\perp x}$ for 
concreteness, since they do not change our conclusion. 

For the equations to be solved consistently it must be assumed that
any commutators between the linear operators can be written as linear
combinations of themselves when they operate on an arbitrary function
on the configuration space. Equivalently they must generate a Lie
algebra with respect to the commutators. In this paper we adopt the
following algebra of commutators between the differential operators.

\begin{eqnarray}
 \left[\hat{\H}_{\perp x},\hat{\H}_{\perp x'}\right]\Psi & = & 
	i\hbar c_{xx'}^{(1)I}\hat{\H}_I\Psi \ \ ,
					\label{eqn:q-algebra1}\\
 \left[\hat{\H}_I,\hat{\H}_{\perp x}\right]\Psi & = & 
	i\hbar c_{Ix}^{(2)x'}\hat{\H}_{\perp x'}\Psi \ \ , 
					\label{eqn:q-algebra2}\\
 \left[\hat{\H}_I,\hat{\H}_J\right]\Psi & = & 
	i\hbar c_{IJ}^{(3)K}\hat{\H}_K\Psi	\label{eqn:q-algebra3}
\end{eqnarray}
for an arbitrary function $\Psi$ on $\M$. Note that the right hand
side of (\ref{eqn:Dirac-algebra1}), (\ref{eqn:Dirac-algebra2}) and
(\ref{eqn:Dirac-algebra3}) are linear in $\hat{\H}_i\Psi$'s,
$\hat{\H}_{\perp}\Psi$'s and  $\hat{\H}_i\Psi$'s, respectively while
the right hand side of (\ref{eqn:q-algebra1}), (\ref{eqn:q-algebra2})
and (\ref{eqn:q-algebra3}) are linear in $\hat{\H}_I\Psi$'s,
$\hat{\H}_{\perp x}\Psi$'s and $\hat{\H}_I\Psi$'s, respectively. Hence 
the algebra (\ref{eqn:q-algebra1}-\ref{eqn:q-algebra3}) is a
generalization of the discretized version of the Dirac algebra
(\ref{eqn:Dirac-algebra1}-\ref{eqn:Dirac-algebra3}). We include the
case when the 'structure constants' $c^{(1)}$, $c^{(2)}$ and $c^{(3)}$
depend on a position in the configuration space $\M$. Although
confirmation of the algebra requires a knowledge of short distance
behaviors of the theory, we simply assume that the algebra does
hold. After detailed investigation of the discretized system, we will
extract results independent of not only the operator ordering but also
a way of the discretization. Note that since a dicretized version of
the Dirac algebra (\ref{eqn:Dirac-algebra1}-\ref{eqn:Dirac-algebra3})
can be written in the form
(\ref{eqn:q-algebra1}-\ref{eqn:q-algebra3}), the system we consider
includes the discretized Wheeler-DeWitt equation as an important
example. Now, for later conveniences we rewrite the assumption
(\ref{eqn:q-algebra1}-\ref{eqn:q-algebra3}) as follows. It is
equivalent to the following set of equalities 
\begin{eqnarray}
 G_x^{\alpha\beta}\partial_{\beta}V_{x'} - 
	G_{x'}^{\alpha\beta}\partial_{\beta}V_x 
	& = & -c^{(1)I}_{xx'}v_I^{\alpha}\ \ ,
				\label{eqn:0th-algebra1}\\
 G_x^{\alpha\gamma}D_{\gamma}v_I^{\beta} +
	G_x^{\beta\gamma}D_{\gamma}v_I^{\alpha}
	& = & c^{(2)x'}_{Ix}G_{x'}^{\alpha\beta}\ \ ,
				\label{eqn:0th-algebra2}\\
 v_I^{\alpha}\partial_{\alpha}V_x 
	& = & -c^{(2)x'}_{Ix}V_{x'}\ \ ,
				\label{eqn:0th-algebra3}\\
 \left[ v_I,v_J\right]^{\alpha}
	& = & -c^{(3)K}_{IJ}v_K^{\alpha} 
				\label{eqn:0th-algebra4}
\end{eqnarray}
and
\begin{eqnarray}
 G_x^{\alpha\beta}D_{\alpha}D_{\beta}V_{x'} -
	G_{x'}^{\alpha\beta}D_{\alpha}D_{\beta}V_x 
	& = & 0\ \ ,	\label{eqn:1st-algebra1}\\
 G_x^{\alpha\beta}D_{\alpha}D_{\beta}v_I^{\gamma} 
	& = & -G_x^{\alpha\beta}v_I^{\delta}
	R^{\gamma}_{\ \alpha\delta\beta}\ \ ,\label{eqn:1st-algebra2}
\end{eqnarray}
where $R$ is a curvature tensor of $D$:
\begin{equation}
 R(X,Y)Z = D_XD_YZ-D_YD_XZ-D_{[X,Y]}Z\ \ .
\end{equation}
Note that, while (\ref{eqn:1st-algebra1}-\ref{eqn:1st-algebra2}) are
affected by the operator ordering,
(\ref{eqn:0th-algebra1}-\ref{eqn:0th-algebra4}) are not. In particular
the fact that $\{ v_I^{\alpha}\}$ is integrable is independent of the
operator ordering since it is a consequence of
(\ref{eqn:0th-algebra4}). We denote the integral surface $G$. In this
paper the quotient space $\M/G$ plays a important role since the
second equation of (\ref{eqn:constraint-eq}) shows that $G$ has
nothing to do with physical degrees of freedom. In the case of the
usual Wheeler-DeWitt equation the quotient space $\M/G$ is called a
superspace. Hence we call the quotient space a {\it superspace} in our 
case, too. Evidently, the superspace is a space of all physically
distinct configurations. 

Next we investigate a classical mechanical system whose quantum
version corresponds to the equations (\ref{eqn:constraint-eq}). Let us
consider a phase space $(T^*\M,dp_{\alpha}\wedge dq^{\alpha})$ and a
Hamiltonian of the form~\footnote{
Hereafter we apply the Einstein's summation rule unless otherwise stated.}
\begin{equation}
 H = \Lambda_{\perp}^x\H_{\perp x} + \Lambda^I\H_I\ \ ,
	\label{eqn:classical-H}
\end{equation}
where $\Lambda_{\perp}^x$ $(\ne 0)$ and $\Lambda^I$ are Lagrange's
multipliers and  
\begin{eqnarray}
 \H_{\perp x} & \equiv & 
	\frac{1}{2}G_x^{\alpha\beta}p_{\alpha}p_{\beta} + 
	V_x\ \ , \nonumber\\
 \H_I & \equiv & 
	v_I^{\alpha}p_{\alpha}. 
\end{eqnarray}
In the expression, $p_{\alpha}$ $(\in T^*\M)$ is a momentum conjugate
to a coordinate $q^{\alpha}$ of
$\M$. (\ref{eqn:0th-algebra1}-\ref{eqn:0th-algebra4}) is equivalent to
the following algebra of Poisson brackets. 
\begin{eqnarray}
 \left\{ \H_{\perp x}, \H_{\perp x'}\right\}_P
	& = & c^{(1)I}_{xx'}\H_I\ \ ,	\nonumber\\
 \left\{ \H_I, \H_{\perp x}\right\}_P
	& = & c^{(2)x'}_{Ix}\H_{\perp x'}\ \ ,\nonumber\\
 \left\{ \H_I, \H_J\right\}_P
	& = & c^{(3)K}_{IJ}\H_K\ \ .	\label{eqn:c-algebra}
\end{eqnarray}
The equivalence is consistent with the well-known fact that the operator
ordering does not affect the corresponding classical dynamics. The
classical equations of motion in the configuration space $\M$ is
obtained from the constraint equations and the Hamilton's equation as
follows~\footnote{
Here the Einstein's summation rule is not applied with respect to
$x$.}. 
\begin{eqnarray}
 \frac{1}{2}G_{x\alpha\beta}\tilde{\N}^{\alpha}\tilde{\N}^{\beta} +
	\left( \Lambda_{\perp}^x\right)^2V_x & = & 0\ \ , \nonumber\\ 
 v_I^{\alpha}\G_{\alpha\beta}\tilde{\N}^{\beta} & = & 0\ \ ,\nonumber\\
 \tilde{\N}^{\beta}\D_{\beta}\tilde{\N}^{\alpha}
 & = & 
	- \G^{\alpha\beta}\partial_{\beta}\V 
	+ \G^{\alpha\beta}\v^{\gamma}\left(
	\partial_{\beta}w_{\gamma}-\partial_{\gamma}w_{\beta}
	\right) \ \ ,	\label{eqn:classical-eq}
\end{eqnarray}
where 
\begin{eqnarray}
 \tilde{\N}^{\alpha} & \equiv & 
	\frac{dq^{\alpha}}{dt}-\v^{\alpha}\ \ ,	\nonumber\\
 \v^{\alpha} & \equiv & \Lambda^I v_I^{\alpha}\ \ ,\nonumber\\
 \G^{\alpha\beta} & \equiv & 
	\Lambda_{\perp}^xG_{x}^{\alpha\beta}\ \ ,\nonumber\\
 \G_{\alpha\beta} & \equiv & 
	\left(\G^{-1}\right)_{\alpha\beta}\ \ ,\nonumber\\
 \V & \equiv & \Lambda_{\perp}^xV_x\ \ ,
\end{eqnarray}
and 
\begin{equation}
 w_{\alpha} \equiv \G_{\alpha\beta}\tilde{\N}^{\beta}\ \ .
\end{equation}
In the expression $\D$ is a covariant derivative compatible with the
metric $\G_{\alpha\beta}$. 

Now we return to the problem of solving the simultaneous differential
equations (\ref{eqn:constraint-eq}) to obtain a wave function. To
extract those properties of the equations which are independent of the
operator ordering we adopt the WKB method. First, without loss of
generality, we can expand the wave function $\Psi$ as 
\begin{equation}
 \Psi = \exp \left[-\frac{1}{\hbar}
	(W^{(0)}+\hbar W^{(1)} + \cdots )\right]\ \ ,
		\label{eqn:WKB-ansatz}
\end{equation}
where $W^{(0)}$, $W^{(1)}$, $\cdots$ are complex functions on
$\M$. Next we solve the differential equations order by order in
$\hbar$, considering $\hbar$ as a small parameter.

\subsection{Lowest-order WKB wave function}

To the lowest order in $\hbar$ the differential equation
(\ref{eqn:constraint-eq}) is reduced to the following simultaneous
differential equations for $W^{(0)}$. 
\begin{eqnarray}
 \frac{1}{2}G_x^{\alpha\beta}
	\partial_{\alpha}W^{(0)}\partial_{\beta}W^{(0)} & = & 
	V_x\ \ , \nonumber\\
 v_I^{\alpha}\partial_{\alpha}W^{(0)} & = & 0\ \ .\label{eqn:lowst}
\end{eqnarray}
These determines $W^{(0)}$ as a function on $\M$, provided that a
suitable boundary condition is attached. When $W^{(0)}$ is real in
a region of the configuration space, we call the corresponding wave
function $\Psi$ an {\it Euclidean wave function}. We call the region
an {\it Euclidean region}. In the remaining of this paper we
investigate the Euclidean wave function. Physical meaning of the
Euclidean wave function will become clear in the following arguments. 

For the real function $W^{(0)}$ we can define a family of vector
fields $\{N_x^{\alpha}\}$ parameterized by the integer $x$: 
\begin{equation}
 N_x^{\alpha} \equiv G_x^{\alpha\beta}\partial_{\beta}W^{(0)}.
\end{equation}
Operating the differential operator
$G_{x'}^{\alpha\beta}D_{\beta}$ on the first equation of
(\ref{eqn:lowst}), we obtain the following equation.
\begin{equation}
 N_x^{\beta}D_{\beta}N_{x'}^{\alpha} = 
	G_{x'}^{\alpha\beta}\partial_{\beta}V_x\ \ , \label{eqn:EOM}
\end{equation}
where we have used the fact that the covariant derivative $D$ is
torsion free and compatible with $G_{x\alpha\beta}$. Note that the
compatibility with $G_{x\alpha\beta}$ is a result of the direct-sum
structure (\ref{eqn:direct-sum}) of $(\M,G_{\alpha\beta})$. 

Since the first of (\ref{eqn:lowst}) is of the form of the
Hamilton-Jacobi equation, we may regard (\ref{eqn:EOM}) as the
corresponding equations of motion for the system. In fact, if we
define a vector field $\N^{\alpha}$ by 
\begin{equation}
 \N^{\alpha} \equiv \Lambda_{\perp}^xN_x^{\alpha}\ \ ,
			\label{eqn:def-N}
\end{equation}
then we can show that 
\begin{eqnarray}
 \frac{1}{2}G_{x\alpha\beta}\N^{\alpha}\N^{\beta} -
	\left( \Lambda_{\perp}^x\right)^2V_x & = & 0\ \ , \nonumber\\ 
 v_I^{\alpha}\G_{\alpha\beta}\N^{\beta} & = & 0\ \ ,\nonumber\\
 \N^{\beta}\D_{\beta}\N^{\alpha}
 & = & 
	\G^{\alpha\beta}\partial_{\beta}\V\ \ .
\end{eqnarray}
Since
$\partial_{\beta}\omega_{\gamma}-\partial_{\gamma}\omega_{\beta}=0$ in 
this case, these equations show that $\N^{\alpha}$ satisfies the
classical equations of motion (\ref{eqn:classical-eq}) in which $V_x$ 
is replaced by $-V_x$, where
$\omega_{\alpha}\equiv\G_{\alpha\beta}\N^{\beta}$. Therefore
$\N^{\alpha}$ defined by (\ref{eqn:def-N}) is a tangent vector field
of a congruence of instantons. In this sense the Euclidean wave
function define a congruence of instantons. Conversely, if a
congruence of instantons satisfies the condition
$\v^{\gamma}(\partial_{\beta}w_{\gamma}-\partial_{\gamma}w_{\beta})=0$, 
then we can construct the corresponding lowest-order Euclidean wave
function as follows. Introduce a parameter $\tau$ by  
\begin{equation}
 \left(\frac{\partial}{\partial\tau}\right)^{\alpha} 
	= \tilde{\N}^{\alpha}\ \ ,
\end{equation}
and
\begin{equation}
 \frac{\partial W^{(0)}}{\partial\tau}=2\V\ \ .
\end{equation}

\subsection{Expansion of the Euclidean wave function around an
instanton}  	\label{subsec:expansion} 

In this subsection we consider a real solution of (\ref{eqn:lowst})
and expand it around an instanton. An expansion of the corresponding
Euclidean wave function is obtained from that.

The second of the constraint equations
(\ref{eqn:constraint-eq}) or the second of (\ref{eqn:lowst})
suggests that any equations for physical degrees of freedom can be
written in forms invariant under any diffeomorphism of $G$. Hence we
can expect that there is a set of real functions
$\{\lambda^x_{\perp}\}$ such that a weighted metric
$\lambda^x_{\perp}G_x^{\alpha\beta}$ and a weighted 'potential'
$\lambda^x_{\perp}V_x$ are invariant under the diffeomorphism. We
first investigate the weighted metric. With the help of
(\ref{eqn:0th-algebra2}) it can be shown that the condition
$\L_{v_I}(\lambda^x_{\perp}G_x^{\alpha\beta})=0$ is equivalent to 
\begin{equation}
 v_I^{\alpha}\partial_{\alpha}\lambda^x_{\perp} = 
	\lambda^{x'}_{\perp}c^{(2)x}_{Ix'}\ \ , 
			\label{eqn:G-inv-lambda}
\end{equation}
where $\L$ represents a Lie derivative in the configuration space $\M$. The invariance of the weighted 
'potential' $\L_{v_I}(\lambda^x_{\perp}V_x)=0$ is equivalent to
(\ref{eqn:G-inv-lambda}), too. Therefore both of
$\lambda^x_{\perp}G_x^{\alpha\beta}$ and$\lambda^x_{\perp}V_x$ are
invariant under the diffeomorphism if and only if
$\{\lambda^x_{\perp}\}$ satisfies (\ref{eqn:G-inv-lambda}). What is
the meaning of the condition (\ref{eqn:G-inv-lambda})?
$c^{(2)x}_{Ix'}$ is the 'structure constant' which appears in the
algebra (\ref{eqn:q-algebra1}-\ref{eqn:q-algebra3}) and represents
a way of transformation of the constraint $\hat{\H}_{\perp x}$ under a
group generated by $\{\hat{\H}_I\}$, which corresponds to $G$. Hence
the condition (\ref{eqn:G-inv-lambda}) means that the coefficient 
$\lambda^x_{\perp}$ must change covariantly under the group
transformation. In order for the equation (\ref{eqn:G-inv-lambda}) to
be solved consistently, the following integrability condition must be
satisfied:
\begin{equation}
 c_{IJ}^{(3)K}c_{Kx'}^{(2)x} + 
	c_{Ix'}^{(2)x''}c_{Jx''}^{(2)x} -
	c_{Jx'}^{(2)x''}c_{Ix''}^{(2)x} +
	v_I^{\alpha}\partial_{\alpha}c_{Jx'}^{(2)x} -
	v_J^{\alpha}\partial_{\alpha}c_{Ix'}^{(2)x} =0\ \ .
		\label{eqn:integrability}
\end{equation}
It can be easily confirmed that (\ref{eqn:integrability}) is actually
satisfied as a consequence of (\ref{eqn:0th-algebra3}) and
(\ref{eqn:0th-algebra4})~\footnote{
The consistency condition (\ref{eqn:integrability}) can be understood
as a consequence of Jacobi identities derived from the commutators 
(\ref{eqn:q-algebra1}-\ref{eqn:q-algebra3}).}. Let $\{\lambda^{\ \
x}_{\perp x'}\}$ be such a 
complete set of linearly independent real solutions of
(\ref{eqn:G-inv-lambda}) that  
\begin{equation} 
 \left.\lambda^{\ \ x}_{\perp x'}
	\right|_{\xi^{\tilde{I}}=\xi_0^{\tilde{I}}} =  
	\delta^x_{x'}\ \ ,
\end{equation}
where $\{\xi^{\tilde{I}}\}$ is a coordinate system of $G$ and
$\{\xi_0^{\tilde{I}}\}$ is a set of constants. Then a general positive 
solution of (\ref{eqn:G-inv-lambda}) can be written as a linear
combination of these solutions:
\begin{equation}
 \lambda^x_{\perp} = 
	\bar{\Lambda}^{x'}_{\perp}\lambda^{\ \ x}_{\perp x'}\ \ ,
		\label{eqn:gen-lambda}
\end{equation}
where $\{\bar{\Lambda}^{x}_{\perp}\}$ is a set of positive functions
on the superspace $\M/G$ in the sense that it is a set of positive
functions on $\M$ satisfying
$\L_{v_I}\bar{\Lambda}^{x}_{\perp}=0$. In the remaining of this
subsection we restrict $\lambda^x_{\perp}$ to this form.

From the set of coefficients $\{\lambda^{\ \ x}_{\perp x'}\}$ we can
define a set of 'metric' $\{{G'}_x^{\alpha\beta}\}$ and a set of
'potentials' $\{\bar{V}_x\}$ by 
\begin{eqnarray}
 {G'}_x^{\alpha\beta} & \equiv & 
	\lambda^{\ \ x'}_{\perp x}G_{x'}^{\alpha\beta}\ \ ,\nonumber\\ 
 \bar{V}_x & \equiv & \lambda^{\ \ x'}_{\perp x}V_{x'}\ \ .
\end{eqnarray}
Here both of $\bar{V}_x$ and ${G'}_x^{\alpha\beta}$ are invariant
under the group transformation of $G$. However, since
${G'}_x^{\alpha\beta}$ may have components in the direction of $G$, it
can not be regarded as a tensor on the superspace $\M/G$ without any
modification, while $\bar{V}_x$ can be. We want to modify it in order
to regard it as a tensor field on $\M/G$. For this purpose define the
following 'projection operators'.
\begin{eqnarray}
 \bar{\G}^{\alpha\beta} & \equiv & 
	{\G'}^{\alpha\beta} -
	v_I^{\alpha}\gamma^{IJ}v_J^{\beta}\ \ ,\nonumber\\
 \bar{\G}_{\alpha\beta} & \equiv & 
	{\G'}_{\alpha\mu}\bar{\G}^{\mu\nu}{\G'}_{\nu\beta}\ \ ,
						\nonumber\\
 \bar{\G}^{\alpha}_{\ \beta} & \equiv &
	\bar{\G}^{\alpha\nu}{\G'}_{\nu\beta}\ \ ,\nonumber\\
 \bar{\G}_{\alpha}^{\ \beta} & \equiv &
	{\G'}_{\alpha\mu}\bar{\G}^{\mu\beta}\ \ ,
\end{eqnarray}
where 
\begin{eqnarray}
 {\G'}^{\alpha\beta} & \equiv & 
	\bar{\Lambda}^{x}_{\perp}{G'}_x^{\alpha\beta}\ \ ,\nonumber\\
 {\G'}_{\alpha\beta} & \equiv & 
	\left({\G'}^{-1}\right)_{\alpha\beta}\ \ ,\nonumber\\
 \gamma_{IJ} & \equiv  &
	{\G'}_{\alpha\beta}v_I^{\alpha}v_J^{\beta}\ \ ,\nonumber\\
 \gamma^{IJ} & \equiv & 
	\left(\gamma^{-1}\right)^{IJ}\ \ .
\end{eqnarray}
It can be easily confirmed by using (\ref{eqn:0th-algebra4}) that 
\begin{equation}
 \L_{v_I}\bar{\G}^{\alpha\beta} = 
	\L_{v_I}\bar{\G}_{\alpha\beta} = 
	\L_{v_I}\bar{\G}^{\alpha}_{\ \beta} = 
	\L_{v_I}\bar{\G}_{\alpha}^{\ \beta} = 0\ \ .
\end{equation}
Then define the modified tensor $\bar{G}_x^{\alpha\beta}$ by
\begin{equation}
 \bar{G}_x^{\alpha\beta} \equiv 
	\bar{\G}^{\alpha}_{\ \mu}{G'}_x^{\mu\nu}
	\bar{\G}_{\nu}^{\ \beta}\ \ .
\end{equation}
It is evident that $\bar{G}_x^{\alpha\beta}$ can be regarded as a
tensor field on the superspace $\M/G$ since
$\L_{v_I}\bar{G}_x^{\alpha\beta}=0$ and
$\bar{G}_x^{\alpha\beta}\bar{\G}_{\beta\gamma}v_I^{\gamma}=0$. Note
that $\bar{\G}^{\alpha\beta}$ equals to a weighted sum of 
$\bar{G}_x^{\alpha\beta}$ as
\begin{equation}
 \bar{\G}^{\alpha\beta} = 
	\bar{\Lambda}^{x}_{\perp}\bar{G}_x^{\alpha\beta}\ \ ,
			\label{eqn:G-Gx-M/G}
\end{equation}
and it can also be regarded as a metric tensor of the superspace $\M/G$. 
In terms of the metric tensor we can rewrite the (\ref{eqn:lowst}) as
the following simultaneous differential equations on $\M/G$ for a real 
function $W^{(0)}$ on $\M/G$.
\begin{equation}
  \frac{1}{2}\bar{G}_x^{\alpha\beta}
	\partial_{\alpha}W^{(0)}\partial_{\beta}W^{(0)} 
	= \bar{V}_x\ \ .\label{eqn:lowst-M/G}
\end{equation}

Our final goal in this subsection is to expand a solution of
(\ref{eqn:lowst-M/G}) around an {\it instanton} in the superspace. For 
this purpose we consider such a congruence of instantons in the
superspace that each instanton is an integral curve of the vector
field $\bar{\N}^{\alpha}$ defined by
\begin{equation}
 \bar{\N}^{\alpha} \equiv 
	\bar{\G}^{\alpha\beta}\partial_{\beta}W^{(0)}\ \ ,
		\label{eqn:N-bar}
\end{equation}
and introduce an Euclidean time $\bar{\tau}$ as a coordinate variable
in $\M/G$ by 
\begin{equation}
 \left(\frac{\partial}{\partial\bar{\tau}}\right)^{\alpha} = 
	\bar{\N}^{\alpha}\ \ .
\end{equation}
If we define a set of vector fields $\{\bar{N}_x^{\alpha}\}$ by 
\begin{equation}
 \bar{N}_x^{\alpha} \equiv 
	\bar{G}_x^{\alpha\beta}\partial_{\beta}W^{(0)}\ \ ,
\end{equation}
then 
\begin{equation}
 \bar{\N}^{\alpha} = \bar{\Lambda}_{\perp}^x \bar{N}_x^{\alpha}\ \ ,
\end{equation}
and $\bar{N}_x^{\alpha}$ generates a time
reparameterization. Commutators between $\bar{N}_x^{\alpha}$ and
$\bar{N}_{x'}^{\alpha}$ is zero by using (\ref{eqn:0th-algebra1}) and
the fact that the covariant derivative $\bar{D}$ is compatible with
$\bar{G}_x^{\alpha\beta}$. Hence the set of vector fields
$\{\bar{N}_x^{\alpha}\}$ is integrable in $\M/G$. Then a coordinate
system $\{\bar{\tau},\eta^{\tilde{x}},\varphi^{\bar{n}}\}$ of $\M/G$
can be introduced so that $\{\bar{\tau},\eta^{\tilde{x}}\}$ is a
coordinate system of the integral surface of
$\{\bar{N}_x^{\alpha}\}$. For these definitions, dependence of
$W^{(0)}$ on $\{\bar{\tau},\eta^{\tilde{x}}\}$ is completely
determined by (\ref{eqn:lowst-M/G}) as
\begin{equation}
 \bar{N}_x^{\alpha}\partial_{\alpha}W^{(0)} = 2\bar{V}_x
		\ \ .\label{W0-G}
\end{equation}
Note that there is a one-to-one map from $\{\varphi^{\bar{n}}\}$ to a
space of all integral surface of the set of vector fields
$\{\bar{N}_x^{\alpha}\}$. Moreover $\{\varphi^{\bar{n}}\}$ is a
maximum set of coordinate variables on which $W^{(0)}$'s dependence is
not completely determined by (\ref{eqn:lowst-M/G}). Thus we can say
that the set of coordinates $\{\varphi^{\bar{n}}\}$ represents all
physical degrees of freedom of perturbations around the instanton. The 
coordinate $\bar{\tau}$ represents an Euclidean time of the congruence 
of instantons, and $\{\eta^{\tilde{x}}\}$ represents degrees of
freedom of time reparameterizations.

Now let us investigate the dependence of $W^{(0)}$ on the physical
coordinates $\{\varphi^{\bar{n}}\}$. For this purpose we derive a
deviation equation of a vector field in Appendix
\ref{app:deviation-eq}. Applying the resulting deviation equation to
the vector fields $\bar{\N}^{\alpha}$ and 
$\bar{Z}_{\bar{n}}^{\alpha}$ 
$\equiv (\partial/\partial\varphi^{\bar{n}})^{\alpha}$
in the superspace $(\M/G,\bar{\G}_{\alpha\beta})$, we obtain
\begin{eqnarray}
 \bar{\D}_F{}_{\perp}\bar{Z}_{\bar{n}}^{\alpha} & = & 
	{}_{\perp}\bar{Z}_{\bar{n}}^{\beta}
	{}_{\perp}\bar{\G}^{\alpha\gamma}
	\bar{\D}_{\gamma}\bar{\D}_{\beta}W^{(0)}\ \ ,\label{eqn:DF-Z}\\
 \bar{\D}_F^2{}_{\perp}\bar{Z}_{\bar{n}}^{\alpha} & = & 
	{}_{\perp}\bar{Z}_{\bar{n}}^{\beta}
	{}_{\perp}\bar{\G}^{\alpha\gamma}
	\left(
	\bar{\D}_{\gamma}\bar{\D}_{\beta}\bar{\V}
	-
	\bar{\R}_{\gamma\rho\beta\sigma}
	\bar{\N}^{\rho}\bar{\N}^{\sigma}
	-
	\frac{3}{2\bar{\V}}
	\partial_{\gamma}\bar{\V}\partial_{\beta}\bar{\V}
	\right)\ \ ,\label{eqn:DF2-Z}
\end{eqnarray}
where $\bar{\D}$ is a covariant derivative compatible with the metric
$\bar{\G}_{\alpha\beta}$, $\bar{\D}_F$ denotes a Fermi derivative
along $\bar{\N}^{\alpha}$ made from $\bar{\D}$, and 
\begin{eqnarray}
 {}_{\perp}\bar{Z}_{\bar{n}}^{\alpha} & \equiv & 
	\bar{Z}_{\bar{n}}^{\alpha} -
	\left(\bar{\G}(\bar{\N},\bar{Z}_{\bar{n}})
	/\bar{\G}(\bar{\N},\bar{\N})\right)
	\bar{\N}^{\alpha}\ \ ,\nonumber\\
 {}_{\perp}\bar{\G}^{\alpha\beta} & \equiv & 
	\bar{\G}^{\alpha\beta} -
	\bar{\N}^{\alpha}\bar{\N}^{\beta}/
	\bar{\G}(\bar{\N},\bar{\N})\ \ ,\nonumber\\ 
 \bar{\R}(X,Y)Z & = & 
	\bar{\D}_X\bar{\D}_YZ-\bar{\D}_Y\bar{\D}_XZ-\bar{\D}_{[X,Y]}Z\ \ .
\end{eqnarray}
The second equation (\ref{eqn:DF2-Z}) makes it possible to investigate
how the vector field ${}_{\perp}\bar{Z}_{\bar{n}}^{\alpha}$ evolves
along a integral line of $\bar{\N}^{\alpha}$: it corresponds to a
linearized equation of motion for the physical degrees of freedom of
perturbations around an instanton. 
With the help of the evolution equation of perturbations, then, the 
first equation (\ref{eqn:DF-Z}) says that the second derivative of
$W^{(0)}$ in the direction of
$\{{}_{\perp}\bar{Z}_{\bar{n}}^{\alpha}\}$ is determined as follows. 
\begin{equation}
 {}_{\perp}\bar{Z}_{\bar{m}}^{\alpha}
	{}_{\perp}\bar{Z}_{\bar{n}}^{\beta}
	\bar{\D}_{\alpha}\bar{\D}_{\beta}W^{(0)} = 
	{}_{\perp}\bar{\G}_{\alpha\beta}
	{}_{\perp}\bar{Z}_{\bar{m}}^{\alpha}
	\bar{\D}_F{}_{\perp}\bar{Z}_{\bar{n}}^{\beta}\ \ .
		\label{eqn:DDF}
\end{equation}
Using these results we can expand a real solution $W^{(0)}$ of the
guide equation around an instanton. For this purpose let us introduce
a set of new variables $\{{}_{\perp}\varphi^{\bar{n}}\}$, each of
which is an affine length of a geodesic normal to the instanton such
that ${}_{\perp}\varphi^{\bar{n}}=0$ along $\Gamma$ and 
\begin{equation}
 \left(\frac{\partial}{\partial{}_{\perp}\varphi^{\bar{n}}}
	\right)^{\alpha} =
	{}_{\perp}\bar{Z}_{\bar{n}}^{\alpha}
	\ \ \mbox{along}\ \ \Gamma\ \ .
\end{equation}
Here $\Gamma$ denotes the instanton around which we intend to
expand a real solution $W^{(0)}$ of the guide equation. Denote the
value of $\eta^{\tilde{x}}$ along the instanton $\Gamma$ by
$\eta_0^{\tilde{x}}$. Then we can expand $W^{(0)}$ by
${}_{\perp}\varphi^{\bar{n}}$ around the instanton $\Gamma$ as 
follows. 
\begin{equation}
 W^{(0)}(\bar{\tau},\eta_0^{\tilde{x}},{}_{\perp}\varphi^{\bar{n}}) = 
	W^{(0)}(\bar{\tau},\eta_0^{\tilde{x}},0) +
	\frac{1}{2}\Omega_{\alpha\beta}(\bar{\tau})
	{}_{\perp}\bar{Z}^{\alpha}_{\bar{m}}
	{}_{\perp}\bar{Z}^{\beta}_{\bar{n}}
	{}_{\perp}\varphi^{\bar{m}}{}_{\perp}\varphi^{\bar{n}} +
	O\left({}_{\perp}\varphi^3\right)\ \ ,
	\label{eqn:expand-F}
\end{equation}
where the matrix $\Omega_{\alpha\beta}$ is defined by 
\begin{equation}
 \Omega_{\alpha\beta}(\bar{\tau}) \equiv \left.
	{}_{\perp}\bar{\G}_{\alpha\gamma}
	\left({}_{\perp}\bar{Z}^{-1}\right)^{\bar{n}}_{\beta}
	\bar{\D}_F{}_{\perp}\bar{Z}_{\bar{n}}^{\gamma}
	\right|_{\eta^{\tilde{x}}=\eta_0^{\tilde{x}},
		{}_{\perp}\varphi^{\bar{n}}=0}\ \ ,
\end{equation}
and ${}_{\perp}\bar{Z}_{\bar{n}}^{\alpha}$ follows the
evolution equation (\ref{eqn:DF2-Z}). In the expression, 
$\left({}_{\perp}\bar{Z}^{-1}\right)^{\bar{n}}_{\alpha}$ is defined by 
\begin{equation}
 {}_{\perp}\bar{Z}_{\bar{n}}^{\alpha}
	\left({}_{\perp}\bar{Z}^{-1}\right)^{\bar{m}}_{\alpha} =
	\delta^{\bar{m}}_{\bar{n}}\ \ ,
\end{equation}
and so that 
$\left({}_{\perp}\bar{Z}^{-1}\right)^{\bar{n}}_{\alpha}
{}_{\perp}\bar{Z}_{\bar{n}}^{\beta}$ is a projection operator to a
subspace generated by $\{{}_{\perp}\bar{Z}_{\bar{n}}^{\alpha}\}$.

As a result, a general real solution of (\ref{eqn:lowst-M/G}) can be
expanded as follows. 
\begin{equation}
 W^{(0)}(\bar{\tau},\eta_0^{\tilde{x}},{}_{\perp}\phi^{\alpha}) 
 =	W^{(0)}(\bar{\tau},\eta_0^{\tilde{x}},0) +
	\frac{1}{2}\Omega_{\alpha\beta}(\bar{\tau})
	{}_{\perp}\phi^{\alpha}{}_{\perp}\phi^{\beta} +
	O\left({}_{\perp}\phi^3\right)\ \ ,
	\label{eqn:expansion}
\end{equation}
where the new variables $\{{}_{\perp}\phi^{\alpha}\}$ are defined by 
\begin{equation}
 {}_{\perp}\phi^{\alpha} \equiv 
	{}_{\perp}\bar{Z}^{\alpha}_{\bar{n}}
	{}_{\perp}\varphi^{\bar{n}}\ \ ,
\end{equation}
provided that its dependence on $\{\bar{\tau},\eta^{\tilde{x}}\}$ is
fixed by (\ref{W0-G}). Note that the set of variables 
$\{{}_{\perp}\phi^{\alpha}\}$ represents all physical degrees of
freedom of perturbations around the instanton. From the expansion
(\ref{eqn:expansion}) we obtain the corresponding expansion of the
Euclidean wave function.
\begin{equation}
 \Psi (\bar{\tau},\eta_0^{\tilde{x}},{}_{\perp}\phi^{\alpha}) 
	= \Psi_0(\bar{\tau})\exp\left[-\frac{1}{\hbar}\left\{
	\frac{1}{2}
	\Omega_{\alpha\beta}(\bar{\tau})
	{}_{\perp}\phi^{\alpha}{}_{\perp}\phi^{\beta} +
	O\left({}_{\perp}\phi^{\alpha 3},
	{}_{\perp}\phi^{\alpha}\hbar\right)\right\}
	\right]\ \ ,
\end{equation}

We have expanded a general real solution of (\ref{eqn:lowst-M/G})
around an instanton $\Gamma$. Finally let us restrict a class of
instantons by imposing a specific boundary condition on the expanded
wave function. We impose the condition that the matrix $\Omega$ is
positive definite in the subspace generated by
$\{{}_{\perp}\bar{Z}_{\bar{n}}^{\alpha}\}$ along the instanton
$\Gamma$ until a possible turning point, where a {\it turning point}
means a point on $\Gamma$ 
at which $\bar{\V}$ becomes zero. The positivity of the matrix means
that probability falls off leaving from the instanton to all physical
directions which are orthogonal to the instanton. It seems satisfied
if the instanton corresponds to a so-called most probable escape path
(MPEP)~\cite{MPEP}. In general, if any first class constraints exist
for a system, they associate with symmetries of the system and there
arises a zero mode problem. However, since by definition $\Omega$ is a
matrix on a linear space of only physical degrees of freedom, $\Omega$
has no zero modes which associate with the first class constraints
(\ref{eqn:constraint-eq}). Thus the positivity is necessary for the
instanton to have a meaning of a MPEP. Note that we can attach this
condition for only a class of instantons, while the expansion
(\ref{eqn:expansion}) is possible around an arbitrary instanton. That
is because the condition on $\Omega$ may restrict behaviors of the
corresponding wave function and may violate some physical criterion
(eg. asymptotic behavior, etc.). Hereafter we restrict a space of all
instantons to the class of those for which we can attach the
positivity of $\Omega$, and we adopt the positivity condition. We
derive significant results from this condition in the next section.


\section{Quantum field theory of perturbations} \label{sec:QFT}

In this section let us see how a quantum field theory of
perturbations is derived from the Euclidean wave function expanded in
the previous section. First we discuss a continuation of the wave
function through the turning point. After that we give a
field-theoretical interpretation of the wave function. 

\subsection{Analytic continuation of the Euclidean wave function}
	\label{subsection:continuation}

As shown in the previous section, the Euclidean wave function is
expanded around an instanton $\Gamma$ as 
\[
 \Psi (\bar{\tau},\eta_0^{\tilde{x}},{}_{\perp}\phi^{\alpha}) 
	= \Psi_0(\bar{\tau})\exp\left[-\frac{1}{\hbar}\left\{
	\frac{1}{2}
	\Omega_{\alpha\beta}(\bar{\tau})
	{}_{\perp}\phi^{\alpha}{}_{\perp}\phi^{\beta} +
	O\left({}_{\perp}\phi^{\alpha 3},
	{}_{\perp}\phi^{\alpha}\hbar\right)\right\}
	\right]\ \ ,
\]
where ${}_{\perp}\phi^{\alpha}=0$ corresponds to  the instanton
$\Gamma$. If we regard ${}_{\perp}\phi^{\alpha}$ as a quantity of
order $O(\hbar^p)$ ($0<p<1$), then we can neglect the terms of order
$O({}_{\perp}\phi^{\alpha 3},{}_{\perp}\phi^{\alpha}\hbar)$ as a
consistent approximation:
\begin{equation}
 \Psi (\bar{\tau},\eta_0^{\tilde{x}},{}_{\perp}\phi^{\alpha})
	= \Psi_0(\bar{\tau})\exp\left[-\frac{1}{2\hbar}
	\Omega_{\alpha\beta}(\bar{\tau})
	{}_{\perp}\phi^{\alpha}{}_{\perp}\phi^{\beta}
	\right]\ \ . \label{eqn:Euclidean}
\end{equation}

It is well known that near a turning point the WKB method is not good. 
So we need a matching condition of WKB wave functions at the turning
point. It is one of the most difficult problems in the WKB approach to 
the Wheeler-DeWitt equation. The difficulty is mainly due to the
following two facts: (1) the configuration space $\M$ is
multi-dimensional (infinite dimensional in a continuous limit); (2)
the `super metric' $\bar{\G}_{\alpha\beta}$ is not positive
definite in the subspace generated by
$\{{}_{\perp}\bar{Z}_{\bar{n}}^{\alpha}\}$.  Vachaspati and Vilenkin
\cite{Vachaspati&Vilenkin} attacked 
the problem by using a simple model and neglecting the second
difficulty. They investigated a Schrodinger equation in a two
dimensional configuration space with a positive definite `super
metric' and obtain the following matching condition of lowest-order
WKB wave functions for the tunneling boundary condition
\cite{Vilenkin}: if the exponent of the wave function is expanded
around a classical solution to second order in perturbations
orthogonal to it, then the wave function of perturbations is
continuous at a turning point. Moreover their result insist that a
proper WKB wave function beyond the turning point is obtained by an
analytic continuation. As a result the matching problem for their
system is equivalent to one for a one-dimensional quantum
system. Returning to our system, if the weighted 'potential'
$\bar{\V}$ does not depend on $\varphi^{\bar{n}}$ near the turning
point with an enough accuracy, then the matching problem for our
system seems equivalent to one for a one-dimensional quantum
system. Hence, in this case, we can expect that their procedure does 
work. Then we obtain the following form of the wave function beyond
the turning point by the analytic continuation with $\bar{\tau}\to
i\bar{t}$. 
\begin{equation}
 \Psi = \Psi_0(i\bar{t})\exp\left[-\frac{1}{2\hbar}
	\Omega_{\alpha\beta}(i\bar{t})
	{}_{\perp}\phi^{\alpha}{}_{\perp}\phi^{\beta}
	\right]\ \ , \label{eqn:Lorentzian}
\end{equation}
where the curve ${}_{\perp}\phi^{\alpha}=0$ corresponds to that 
solution of the classical equation of motion (\ref{eqn:classical-eq}) 
which is the analytic continuation of the instanton $\Gamma$, and 
$\{{}_{\perp}\phi^{\alpha}\}$ denotes physical perturbations
around the classical path. Here the matrix $\Omega_{\alpha\beta}$ can
be written as 
\begin{equation}
 \Omega_{\alpha\beta}(i\bar{t})
	= -i {}_{\perp}\bar{\G}_{\alpha\gamma}
	\left({}_{\perp}\bar{\Z}^{-1}\right)^{\bar{n}}_{\beta}
	\bar{\D}_F{}_{\perp}\bar{\Z}_{\bar{n}}^{\gamma}\ \ ,
	\label{eqn:Omega-Lorentz}
\end{equation}
where $\bar{\D}_F$ denotes a Fermi-derivative along the
analytically-continued classical path and
${}_{\perp}\bar{\Z}_{\bar{n}}^{\alpha}$ is the analytic continuation
of ${}_{\perp}\bar{Z}_{\bar{n}}^{\alpha}$.

\subsection{Field-theoretical interpretation}
	\label{subsec:FT-interp}

In quantum field theory, normalized mode functions play a central role. 
The mode functions are normalized in terms of a so-called 
Klein-Gordon inner product. Thus, for our discretized system, we
propose the following definition of an inner product between two
complex vector fields: for complex vector fields $X^{\alpha}$ and
$Y^{\alpha}$ on the superspace $\M/G$, 
\begin{equation}
 (X,Y)_{KG} \equiv 
	-i{}_{\perp}\bar{\G}_{\alpha\beta}\left(
	{}_{\perp}X^{\alpha}\bar{\D}_F{}_{\perp}{Y^*}^{\beta} -
	{}_{\perp}{Y^*}^{\alpha}\bar{\D}_F{}_{\perp}X^{\beta}\right)
		\ \ ,\label{eqn:KG-norm}
\end{equation}
where $*$ denotes a complex conjugation. For this definition, it can
be easily proved that the inner product  between
$\{{}_{\perp}\bar{\Z}_{\bar{n}}^{\alpha}\}$ is constant along  the
classical path:  
\begin{equation}
 \frac{\partial}{\partial\bar{t}}
 ({}_{\perp}\bar{\Z}_{\bar{m}},{}_{\perp}\bar{\Z}_{\bar{n}})_{KG} 
	= 0 \ \ .\label{eqn:const-KG}
\end{equation}

We want to normalize the set of vectors 
$\{{}_{\perp}\bar{\Z}_{\bar{n}}^{\alpha}\}$ since each of them is a 
solution of perturbed equation of motion and corresponds to the mode 
function in a continuous limit. By introducing a complex regular
matrix $\zeta_k^{\bar{n}}$ whose elements are constant, we take linear 
combinations of complex conjugates of
$\{{}_{\perp}\bar{\Z}_{\bar{n}}^{\alpha}\}$ as 
\begin{equation}
 {}_{\perp}u_k^{\alpha} \equiv \zeta_k^{\bar{n}}
	{}_{\perp}\bar{\Z}_{\bar{n}}^{*\alpha}\ \ .
\end{equation}
Evidently ${}_{\perp}u_k^{\alpha}$ is a candidate for the normalized
mode function. By using (\ref{eqn:Omega-Lorentz}) we can express the
inner product between $\{{}_{\perp}u_k^{\alpha}\}$ in terms of $\Omega$
as 
\begin{equation}
 ({}_{\perp}u_k,{}_{\perp}u_{k'})_{KG} = 
	\left[{}_{\perp}u(\Omega +\Omega^{\dagger})
	{}_{\perp}u^{\dagger}\right]_{kk'}\ \ .\label{eqn:uu-KG}
\end{equation}

The positivity of the $\Omega$ in the Euclidean region, which has been
introduced in the last paragraph of subsection~\ref{subsec:expansion},
means that the matrix $({}_{\perp}u_k,{}_{\perp}u_{k'})_{KG}$
calculated in (\ref{eqn:uu-KG}) is
positive definite at the turning point. Hence, due to the constancy of
the matrix elements implied by (\ref{eqn:const-KG}), we can choose the
transformation matrix $\zeta_k^{\bar{n}}$ so that
\begin{equation}
 ({}_{\perp}u_k,{}_{\perp}u_{k'})_{KG} = \delta_{kk'}
			\label{eqn:KG-normalization}
\end{equation}
along the analytically-continued classical path. Then
$\{{}_{\perp}u_{k}^{\alpha}\}$ is specified as a set of 
orthonormalized solutions of the simultaneous equations 
\begin{eqnarray}
 \bar{\D}_F^2{}{}_{\perp}u_{k}^{\alpha} & = &
	- {}_{\perp}u_{k}^{\beta}
	{}_{\perp}\bar{\G}^{\alpha\gamma}
	\left(
	\bar{\D}_{\gamma}\bar{\D}_{\beta}\bar{\V}
	+
	\bar{\R}_{\gamma\rho\beta\sigma}
	\left(\frac{\partial}{\partial\bar{t}}\right)^{\rho}
	\left(\frac{\partial}{\partial\bar{t}}\right)^{\sigma}
	-
	\frac{3}{2\bar{\V}}
	\partial_{\gamma}\bar{\V}\partial_{\beta}\bar{\V}
	\right)\ \ ,\label{eqn:eq-u1}\\
 0 & = & \bar{\G}_{\alpha\beta}\bar{\N}^{\alpha}
	{}_{\perp}u_{k}^{\beta}\ \ ,\label{eqn:eq-u2}
\end{eqnarray}
and is a complete set of basis vectors in a subspace generated by
$\{(\partial /\partial{}_{\perp}\varphi^{\bar{n}})^{\alpha}\}$. 
Equation (\ref{eqn:eq-u1}-\ref{eqn:eq-u2}) is a linearized evolution equation of orthogonal 
perturbations around the analytically-continued classical path. The 
boundary condition of ${}_{\perp}u_{k}^{\alpha}$ is that the matrix
$\Omega_{\alpha\beta}$ defined by 
\begin{equation}
 \Omega_{\alpha\beta} \equiv 
 -i {}_{\perp}\bar{\G}_{\alpha\gamma}
	({}_{\perp}u^{-1})_{\beta}^{*k}
 \bar{\D}_{F}{}_{\perp}u_k^{*\gamma} \label{eqn:Omega-u}
\end{equation}
is real and positive definite in the subspace generated by
$\{{}_{\perp}\bar{Z}_{\bar{n}}^{\alpha}\}$ when it is
analytically-continued back 
to the Euclidean region (see arguments in the last paragraph of
subsection~\ref{subsec:expansion}), where $*$ denotes a complex 
conjugation and $({}_{\perp}u^{-1})_{\alpha}^{k}$ is defined by 
\begin{equation}
 {}_{\perp}u_k^{\alpha}({}_{\perp}u^{-1})_{\alpha}^{k'} =
	\delta_k^{k'}\ \ ,
\end{equation}
and so that $({}_{\perp}u^{-1})_{\alpha}^{k}{}_{\perp}u_k^{\beta}$ is
a projection operator to the subspace generated by 
$\{(\partial /\partial{}_{\perp}\varphi^{\bar{n}})^{\alpha}\}$.

Finally we show that the wave function (\ref{eqn:Lorentzian}) can
actually be interpreted in terms of a quantum-mechanical system, which
will become a quantum field theory in a continuous limit. Consider a
quantum-mechanical system described by the hamiltonian 
\begin{equation}
 \hat{H} = \frac{1}{2}\left(
	{}_{\perp}\bar{\G}^{\alpha\beta}
	{}_{\perp}\hat{\pi}_{\alpha}{}_{\perp}\hat{\pi}_{\beta} + 
	{}_{\perp}V_{\alpha\beta}
	{}_{\perp}\hat{\phi}^{\alpha}{}_{\perp}\hat{\phi}^{\beta}
	\right)	\ \ ,\label{eqn:effective-H}
\end{equation}
and the equation of motion 
\begin{equation}
 i\hbar\bar{\D}_F\hat{O} = \left[\hat{O},\hat{H}\right]
\end{equation}
for an arbitrary operator $\hat{O}$, where
${}_{\perp}\hat{\pi}_{\alpha}$ is a momentum conjugate to 
${}_{\perp}\hat{\phi}^{\alpha}$ and 
\begin{equation}
 {}_{\perp}V_{\alpha\beta} \equiv
	{}_{\perp}\bar{\G}_{\alpha}^{\mu}
	{}_{\perp}\bar{\G}_{\beta}^{\nu}
	\left(
	\bar{\D}_{\mu}\bar{\D}_{\nu}\bar{\V}
	+
	\bar{\R}_{\mu\rho\nu\sigma}
	\left(\frac{\partial}{\partial\bar{t}}\right)^{\rho}
	\left(\frac{\partial}{\partial\bar{t}}\right)^{\sigma}
	-
	\frac{3}{2\bar{\V}}
	\partial_{\mu}\bar{\V}\partial_{\nu}\bar{\V}
	\right)\ \ . \label{eqn:perp_V}
\end{equation}
The equation of motion says that the operator
${}_{\perp}\hat{\phi}^{\alpha}$ can be expanded as follows. 
\begin{equation}
 {}_{\perp}\hat{\phi}^{\alpha} = \hbar^{1/2}\sum_k\left(
	\hat{a}_k{}_{\perp}u_k^{\alpha} + 
	\hat{a}_k^{\dagger}{}_{\perp}u_k^{*\alpha}\right)\ \ ,
	\label{eqn:Hei-phi}
\end{equation}
where $\hat{a}_k$ and $\hat{a}_k^{\dagger}$ are annihilation and
creation operators which satisfy
\begin{eqnarray}
 \left[\hat{a}_k,\hat{a}_{k'}^{\dagger}\right] & = & \delta_{kk'}\ \ , 
	\nonumber \\ 
 \left[\hat{a}_k,\hat{a}_{k'}\right] & = & 0\ \ ,\nonumber\\
 \left[\hat{a}_k^{\dagger},\hat{a}_{k'}^{\dagger}\right] & = & 0 \ \ .
\end{eqnarray}
For the representation (\ref{eqn:Hei-phi}) with the normalization
(\ref{eqn:KG-normalization}), the annihilation operator
$\hat{a}_k$ can be expressed as a linear combination of terms of the
form  
\begin{equation}
 {}_{\perp}\hat{\pi}_{\alpha}-
	i\Omega_{\alpha\beta}{}_{\perp}\hat{\phi}^{\beta}\ \ .
\end{equation}
Hence the wave function (\ref{eqn:Lorentzian}) is annihilated by all
the annihilation operators: 
\begin{equation}
 \hat{a}_k\Psi = 0
\end{equation}
for all $k$. This says that the wave function (\ref{eqn:Lorentzian})
represents the vacuum state determined by the vectors
$\{{}_{\perp}u_{k}^{\alpha}\}$. Thus, in a continuous limit, the
corresponding wave function can be interpreted in terms of a quantum
field theory: it represents the vacuum state determined by that set of 
positive-frequency mode functions which is a continuous limit of 
$\{{}_{\perp}u_{k}^{\alpha}\}$.

\subsection{Reduced Lagrangian}

When we intend to apply the formalism developed in this paper, what we 
have to do is: (1) to fix a MPEP; (2) to give a coordinate system in a
space of all gauge-invariant perturbations (physical perturbations)
around the MPEP; (3) to seek explicit form of the effective
Hamiltonian (\ref{eqn:effective-H}) and the matrix
(\ref{eqn:Omega-u}); (4) to calculate any wanted quantities by using 
the quantum field theory described by the effective Hamiltonian. The
positive-frequency mode function is defined so that the matrix
(\ref{eqn:Omega-u}) is real and positive definite in the subspace
spanned by the gauge-invariant perturbations in the Euclidean
region. 

In this subsection we show that the effective Hamiltonian and
the matrix can be obtained from an reduced Lagrangean of the
gauge-invariant perturbations. Thus the reduced Lagrangean gives a
convenient prescription to perform the procedure (3). 

The classical Hamiltonian (\ref{eqn:classical-H}) corresponds to the
Lagrangian 
\begin{eqnarray}
 L & = & p_{\alpha}\dot{q}^{\alpha} - H \nonumber\\
   & = & \frac{1}{2}\G_{\alpha\beta}(\dot{q}^{\alpha}-\v^{\alpha})
	(\dot{q}^{\beta}-\v^{\beta}) - \V \ \ .
\end{eqnarray}
This Lagrangian describes a dynamics in the full configuration space
$\M$. When we go to the superspace $\M/G$, the corresponding reduced
Lagrangian $\bar{L}$ in $\M/G$ is 
\begin{equation}
 \bar{L} = \frac{1}{2}\bar{\G}_{\alpha\beta}\dot{\bar{q}}^{\alpha}
	\dot{\bar{q}}^{\beta} - \bar{\V} \ \ ,
\end{equation}
where $\{\bar{q}^{\alpha}\}$ is a coordinate system in $\M/G$. 
The kinetic part for the gauge-invariant perturbations is~\footnote{
Note that terms linear in ${}_{\perp}\phi^{\alpha}$ is automatically
zero because of the background equation.
}
\begin{equation}
 \frac{1}{2}{}_{\perp}\bar{\G}_{\alpha\beta}
	{}_{\perp}\dot{\phi}^{\alpha}{}_{\perp}\dot{\phi}^{\beta}
	\ \ (\in\bar{L})\ \ .
\end{equation}
Thus the form of ${}_{\perp}\bar{\G}_{\alpha\beta}$, which is
necessary and sufficient to obtain the explicit form of the matrix
(\ref{eqn:Omega-u}), can be read from the kinetic part of the reduced
Lagrangian. 

When the dot $\dot{}$ is identified with $\bar{\D}_F$, $\bar{L}$ must
derive the field equation (\ref{eqn:eq-u1}) with
${}_{\perp}u_{k}^{\alpha}$ replaced by ${}_{\perp}\phi^{\alpha}$
since by definition the field equation (\ref{eqn:eq-u1}) is nothing
else the evolution equation of the orthogonal perturbations. From this 
fact, the quadratic part of the reduced Lagrangian must be of the form 
\begin{equation}
 \frac{1}{2}{}_{\perp}\bar{\G}_{\alpha\beta}
	{}_{\perp}\dot{\phi}^{\alpha}{}_{\perp}\dot{\phi}^{\beta}
	-\frac{1}{2}{}_{\perp}V_{\alpha\beta}
	{}_{\perp}\phi^{\alpha}{}_{\perp}\phi^{\beta}
	\ \ (\in \bar{L})\ \ ,
\end{equation}
where ${}_{\perp}V_{\alpha\beta}$ is defined by
(\ref{eqn:perp_V}). The effective Hamiltonian (\ref{eqn:effective-H})
can be obtained from the effective action by usual prescription. Thus,
in a continuous limit, the reduced Lagrangian leads us to the
effective quantum field theory considered in the previous subsection.

After all, when we intend to investigate a state of the physical
perturbations after a quantum tunneling, what we have to do is: (a) to 
fix a MPEP; (b) to give a coordinate system in a space of
all gauge-invariant perturbations around the MPEP; (c) to obtain the 
reduced Lagrangian of the gauge-invariant perturbations by simply
reducing the original Lagrangian to the space of all gauge-invariant
perturbations; (d) to use the quantum field theory of the
perturbations constructed from the reduced Lagrangian. From the
effective action we can read off the form of the matrix
(\ref{eqn:Omega-u}), which must be real and positive in the subspace 
generated by gauge-invariant perturbations in the Euclidean region.


\section{Summary and discussion}
	\label{sec:Sammary} 

Throughout this paper, we have
investigated the simultaneous differential equations
(\ref{eqn:constraint-eq}) with the algebra
(\ref{eqn:q-algebra1}-\ref{eqn:q-algebra3}) by
using the WKB method. We have mentioned that, in a continuous limit,
(\ref{eqn:constraint-eq}) and
(\ref{eqn:q-algebra1}-\ref{eqn:q-algebra3}) become 
generalizations of the Wheeler-DeWitt equation (\ref{eqn:WD-eq}) and
the Dirac algebra (\ref{eqn:Dirac-algebra1}-\ref{eqn:Dirac-algebra3}),
respectively. We have 
defined an Euclidean wave function as that solution of the equations
(\ref{eqn:constraint-eq}) whose lowest-order part in the WKB expansion
(\ref{eqn:WKB-ansatz}) is real in a region of the configuration
space. We have called the region an Euclidean region. The lowest-order
part in $\hbar$ has been expanded in the superspace around an
instanton, by using a deviation equation of a vector field tangent to
a congruence of instantons. The instanton around which we expand the
wave function corresponds to a so-called most probable escape path
(MPEP). Then we have shown that, when the expanded wave function is
analytically-continued beyond the Euclidean region, the continued wave
function can be understood in terms of a quantum-mechanical system,
which will become a quantum field theory in a continuous limit. The
corresponding state of physical perturbations around the
analytically-continued classical path is the vacuum state determined
by 'positive-frequency mode functions'
$\{{}_{\perp}u_{k}^{\alpha}\}$. Each 'mode function' must be
normalized with respect to the norm (\ref{eqn:KG-norm}) and satisfy
the evolution equations (\ref{eqn:eq-u1}-\ref{eqn:eq-u2}), which
correspond to 
linearized classical equations of motion for perturbations orthogonal
to the analytically-continued classical path. The positive-frequency
mode functions must satisfy also the following boundary condition: the 
matrix $\Omega_{\alpha\beta}$ defined by (\ref{eqn:Omega-u}) must be
real and positive definite in the subspace generated by
$\{{}_{\perp}\bar{Z}_{\bar{n}}^{\alpha}\}$ when it is
analytically-continued back to 
the Euclidean region. Note that the boundary condition of the mode
functions is dependent of the matching condition of the WKB wave
functions. In subsection \ref{subsection:continuation}, in the case
that the weighted 'potential' $\bar{\V}$ does not depend on
$\varphi^{\bar{n}}$ near the turning point with an enough accuracy,
we have adopted such a matching condition that the WKB wave functions
are analytic continuations of each other. If a future analysis for
more general case gives a different matching condition of the WKB wave
functions, then the mode functions must be matched with those in the
Euclidean region in a different way and the corresponding matrix
$\Omega_{\alpha\beta}$ must be real and positive definite in the
subspace generated by
$\{{}_{\perp}\bar{Z}_{\bar{n}}^{\alpha}\}$. Anyway, the 
state of perturbations after a quantum tunneling is determined
uniquely by the positive-frequency mode functions. Thus a quantum
field theory is effective to investigate a state of physical
perturbations after a quantum tunneling. We have shown that the
effective Lagrangian describing the field theory is obtained by simply
reducing the original Lagrangian to the subspace spanned by the
physical perturbations around the MPEP.

The result of this paper does not depend on the operator ordering
since we have concentrated only on
(\ref{eqn:0th-algebra1}-\ref{eqn:0th-algebra4}) and (\ref{eqn:lowst}), 
which are independent of the operator ordering. Moreover the
result can be applied to a general background, which corresponds to a
most probable escape path (MPEP). Thus, for a general MPEP, a 
quantum field theory is effective to investigate a state of all
physical perturbations after a quantum tunneling with gravity. Due to
the result of this paper we can safely use the quantum field theory to 
obtain a CMB anisotropy, a spectrum of primordial gravitational
waves, etc. in some inflationary scenarios. For example, in the
one-bubble inflationary scenario, Tanaka and Sasaki \cite{Grav-wave}
calculated the reduced Lagrangian and quantized the gravitational
perturbations. Although they did not justify their treatment from the
point of view of the Wheeler-DeWitt equation, the result of this paper 
do justify it without writing down the explicit form of the perturbed 
Wheeler-DeWitt equation.

Finally we propose a possible definition of a two point function to
strengthen the field theoretical interpretation. For a function $f$ on
$\M/G$, define a one-parameter family of expectation values 
$\langle\ f\ \rangle_{\bar{t}}$ along the analytically-continued
classical path. 
\begin{equation}
\langle\ f\ \rangle_{\bar{t}}
	\equiv
	\left[\frac{\int(\prod_{\bar{n}}d{}_{\perp}\varphi^{\bar{n}})
	\sqrt{|\det'{}_{\perp}\bar{\G}|}
	\left|\Psi\right|^2f}
	{\int(\prod_{\bar{n}}d{}_{\perp}\varphi^{\bar{n}})
	\sqrt{|\det'{}_{\perp}\bar{\G}|}
	\left|\Psi\right|^2}
	\right]_{\bar{t},\eta^{\tilde{x}}=\eta_0^{\tilde{x}}}
	\ \ ,
\end{equation}
where $\det'$ represents a determinant of the followed matrix
restricted to the subspace generated by 
$\{(\partial /\partial{}_{\perp}\varphi^{\bar{n}})^{\alpha}\}$. It is
evident that for the wave function (\ref{eqn:Lorentzian}) the
expectation value of ${}_{\perp}\phi^{\alpha}$ is zero: 
$\langle{}_{\perp}\phi^{\alpha}\rangle_{\bar{t}}=0$. 
On the other hand the expectation value of 
$\{{}_{\perp}\phi^{\alpha},{}_{\perp}\phi^{\beta}\}$ 
is not zero: 
\begin{equation}
 \langle\left\{{}_{\perp}\phi^{\alpha},
	{}_{\perp}\phi^{\beta}\right\}
	\rangle_{\bar{t}} \approx
	\hbar\left[ (\Omega+\Omega^{\dagger})^{-1}
	\right]^{\alpha\beta} +
	\hbar\left[ (\Omega+\Omega^{\dagger})^{-1}
	\right]^{\beta\alpha}\ \ ,\label{eqn:lambda-lambda}
\end{equation}
where the right hand side is calculated on the classical path. Note
that '$\approx$' means that the difference between the right and the
left hand side is sufficiently small when the wave function is so
peaked along the classical path that an orthogonal transformation of
the hermite part of $\Omega$ and the evaluation of the expectation
value is approximately commutable. From (\ref{eqn:uu-KG}) we can show
that 
\begin{equation}
 (\Omega +\Omega^{\dagger})^{-1} = 
	{}_{\perp}u^{\dagger}({}_{\perp}u,{}_{\perp}u)_{KG}^{-1}
	{}_{\perp}u\ \ ,
\end{equation}
and 
\begin{equation}
 \langle\left\{{}_{\perp}\phi^{\alpha},
	{}_{\perp}\phi^{\beta}\right\}
	\rangle_{\bar{t}} \approx
	\hbar\left[{}_{\perp}u^{\dagger}
	({}_{\perp}u,{}_{\perp}u)_{KG}^{-1}
	{}_{\perp}u	\right]^{\alpha\beta} +
	\hbar\left[{}_{\perp}u^{\dagger}
	({}_{\perp}u,{}_{\perp}u)_{KG}^{-1}
	{}_{\perp}u	\right]^{\beta\alpha}\ \ .
\end{equation}
In subsection~\ref{subsec:FT-interp} we have shown that the matrix
$\zeta_k^{\bar{n}}$ can be chosen so that 
$({}_{\perp}u_k,{}_{\perp}u_{k'})_{KG}=\delta_{kk'}$. For this choice
of $\zeta_k^{\bar{n}}$, the two point function is reduced to  
\begin{equation}
\langle\left\{{}_{\perp}\phi^{\alpha},{}_{\perp}\phi^{\beta}\right\}
	\rangle_{\bar{t}} \approx
	\hbar\sum_k\left({{}_{\perp}u_{k}^*}^{\alpha}
	{}_{\perp}u_{k}^{\beta} + 
	{}_{\perp}u_{k}^{\alpha}
	{{}_{\perp}u_{k}^*}^{\beta}\right)\ \ . \label{eqn:2point}
\end{equation}
The right hand side of (\ref{eqn:2point}) can be understood as a two 
point function of the quantum-mechanical system introduced in
subsection~\ref{subsec:FT-interp} as follows. The  representation
(\ref{eqn:Hei-phi}) leads the two point function
\begin{equation}
 \langle 0|\{{}_{\perp}\hat{\phi}^{\alpha},
    {}_{\perp}\hat{\phi}^{\beta}\}|0\rangle = \hbar
	\sum_k\left({}_{\perp}u_k^{*\alpha}{}_{\perp}u_k^{\beta} +
	{}_{\perp}u_k^{\alpha}{}_{\perp}u_k^{*\beta}\right)
		\ \ ,\label{eqn:2point'}
\end{equation}
where the state $|0\rangle$ is the vacuum state of the
quantum-mechanical system:
\begin{equation}
 \hat{a}_k|0\rangle = 0 
\end{equation}
for ${}^{\forall}k$. The two point function (\ref{eqn:2point'}) is the
right hand side of (\ref{eqn:2point}) itself. Thus the two point
function based on the wave function (\ref{eqn:Lorentzian}) is
equivalent to that based on the quantum-mechanical system.

\vskip 1cm

\centerline{\bf Acknowledgments}
The author thanks Prof. H. Kodama for continuous encouragement and 
valuable comments. He also thanks Dr. T. Tanaka, Prof. M. Sasaki and 
Y. Mino for helpful discussions.


\appendix

\section{Deviation equation of a vector field}
\label{app:deviation-eq}

In this appendix we derive a deviation equation for a spacelike (and 
timelike) vector field in a pseudo-Riemannian manifold. By spacelike
(timelike) we mean that its norm with respect to the metric is
positive (negative). 

Let $(M,g_{\alpha\beta})$ be a pseudo-Riemannian manifold and
$N^{\alpha}$ be a vector field on $M$. It is assumed that
$g_{\alpha\beta}N^{\alpha}N^{\beta}\ne 0$ in a region $E$ 
$(\in M)$. Then introduce another vector field $Z^{\alpha}$ in $E$
such that  
\begin{equation}
 \left[ N,Z\right]^{\alpha} = 0\ \ .\label{eqn:N-Z}
\end{equation}
Locally this condition is a necessary and sufficient condition in
order for $N^{\alpha}$ and $Z^{\alpha}$ to define independent
coordinate variables by 
\begin{eqnarray}
 \left(\frac{\partial}{\partial\tau}\right)^{\alpha} & = & 
	N^{\alpha}\ \ , \nonumber\\
 \left(\frac{\partial}{\partial\lambda}\right)^{\alpha} & = & 
	Z^{\alpha}\ \ .
\end{eqnarray}
When we apply equations derived in this appendix to the system
investigated in subsection~\ref{subsec:expansion}, $N^{\alpha}$
corresponds to a gradient vector filed (\ref{eqn:N-bar}) 
of a real solution of (\ref{eqn:lowst-M/G}). 
Then $\lambda$ corresponds to a coordinate $\varphi^{\bar{n}}$ 
distinguishing physically different integral lines of $N^{\alpha}$. 
The condition (\ref{eqn:N-Z}) can be written explicitly as 
\begin{equation}
 \dot{Z}^{\alpha} = Z^{\beta}\nabla_{\beta}N^{\alpha}\ \ , 
		\label{eqn:dot-Z}
\end{equation}
where $\nabla$ is a covariant derivative compatible with
$g_{\alpha\beta}$ and 
\begin{equation}
 \dot{X}^{\alpha} \equiv N^{\beta}\nabla_{\beta}X^{\alpha}
\end{equation}
for $X^{\alpha}\in TM$. Operating $N^{\beta}\nabla_{\beta}$ to it, we
obtain the following equation.
\begin{equation}
 \ddot{Z}^{\alpha} = Z^{\beta}\nabla_{\beta}\dot{N}^{\alpha}
	-R^{\alpha}_{\ \rho\beta\sigma}N^{\rho}N^{\sigma}Z^{\beta}
			\ \ ,\label{eqn:ddot-Z}
\end{equation}
where $R(X,Y)Z=\nabla_X\nabla_YZ-\nabla_Y\nabla_XZ-\nabla_{[X,Y]}Z$. 

Without any modification, these equations are not effective enough to 
investigate the dependence of the wave function on the perturbation
around the instanton. It is because $Z^{\alpha}$ may have a component
in the direction of $N^{\alpha}$. We want to modify these equations so 
that $Z^{\alpha}$ appears as a vector projected to a hypersurface
orthogonal to $N^{\alpha}$ in the modified equations. To make the 
modification possible, we define a so-called Fermi transported
basis. First a Fermi derivative $\nabla_F$ along $N^{\alpha}$ is
defined by 
\begin{equation}
 \nabla_FX^{\alpha} = \dot{X}^{\alpha}+\frac{1}{N^2}g(X,\dot{N})N^{\alpha}
	- \frac{1}{N^2}g(X,N)\dot{N}
\end{equation}
for $X^{\alpha}\in TM$, where $N^2\equiv
g(N,N)$~\cite{Hawking&Ellis}. Note that $N^2$ is positive (negative)
when $N^{\alpha}$ is spacelike (timelike). The Fermi derivative has
the following well known properties. 
\begin{enumerate}
 \item If $N^{\alpha}$ is tangent to a congruence of geodesics, then
	$\nabla_FX^{\alpha}=\dot{X}^{\alpha}$ for $X^{\alpha}\in TM$.
 \item $\nabla_F{}_{\perp}X^{\alpha} = 
	 \ _{\perp}(\ _{\perp}X)\dot{}\ ^{\alpha}$ for $X^{\alpha}\in
	TM$, where  
	$\ _{\perp}X^{\alpha}\equiv 
	X^{\alpha}-N^{\alpha}g(N,X)/N^2$.
 \item $\nabla_F(N^{\alpha}/\sqrt{|N^2|})=0$.
 \item If $X^{\alpha}$ and $Y^{\alpha}$ $(\in TM)$ satisfy
	$\nabla_FX^{\alpha}=\nabla_FY^{\alpha}=0$, then
	$N^{\alpha}\partial_{\alpha}g(X,Y)=0$. 
\end{enumerate}
The first and the second properties suggest that the Fermi derivative
is effective for our purpose. From the third and the forth properties, 
along an integral curve of $N^{\alpha}$ the Fermi transported basis
can be defined by
\begin{eqnarray}
 e_0^{\alpha} & \equiv & N^{\alpha}/\sqrt{|N^2|}\ \ ,\nonumber\\
 \nabla_Fe_i^{\alpha} & = & 0\ \ ,
\end{eqnarray}
and so that $g(e_0,e_i)=0$ and $g(e_i,e_j)=\eta_{ij}$, where
$\eta_{ij}$ is a constant regular matrix. By using this basis, we can
expand $Z^{\alpha}$ as 
\begin{equation}
 Z^{\alpha} = Z^0e_0^{\alpha} + Z^ie_i^{\alpha}\ \ ,\label{eqn:Z-e}
\end{equation}
where $Z^0\equiv g(Z,e_0)/g(e_0,e_0)$ and
$Z^i\equiv(\eta^{-1})^{ij}g(Z,e_j)$. Substituting (\ref{eqn:Z-e}) into 
(\ref{eqn:dot-Z}) and (\ref{eqn:ddot-Z}), we can show by explicit
calculation that 
\begin{eqnarray}
 \frac{\partial}{\partial\tau}Z^i & = & 
 	Z^j\nabla_jN^i\ \ ,\nonumber\\
 \frac{\partial^2}{\partial\tau^2}Z^i & = & 
 	Z^j\nabla_j\dot{N}^i - 
	R^i_{\ \rho j\sigma}N^{\rho}N^{\sigma}Z^j 
	- \frac{1}{N^2}\dot{N}^iZ^j\eta_{jk}\dot{N}^k
	- \frac{1}{N^2}\dot{N}^iZ^j\nabla_jN^2\ \ ,
\end{eqnarray}
where
\begin{eqnarray}
 \dot{N}^i & \equiv & (\eta^{-1})^{ik}g_{\alpha\beta}
 	e_{k}^{\alpha}\dot{N}^{\beta}\ \ ,\nonumber\\
 \nabla_jN^i & \equiv & (\eta^{-1})^{ik}g_{\alpha\beta}
 	e_{k}^{\alpha}e_{j}^{\gamma}\nabla_{\gamma}N^{\beta}
				\ \ ,\nonumber\\
 \nabla_j\dot{N}^i & \equiv & (\eta^{-1})^{ik}g_{\alpha\beta}
 	e_{k}^{\alpha}e_{j}^{\gamma}\nabla_{\gamma}\dot{N}^{\beta}
				\ \ ,\nonumber\\
 R^i_{\ \rho j\sigma} & \equiv & (\eta^{-1})^{ik}g_{\alpha\beta}
 	e_{k}^{\alpha}e_{j}^{\gamma}R^{\beta}_{\ \rho\gamma\sigma}
				\ \ ,\nonumber\\
 \nabla_j(N^2) & \equiv & e_j^{\alpha}\partial_{\alpha}(N^2)\ \ .
\end{eqnarray}
The result can be rewritten as a covariant form:
\begin{eqnarray}
 \nabla_F{}_{\perp}Z^{\alpha} & = & 
	{}_{\perp}Z^{\beta}
	{}_{\perp}(\nabla_{\beta}N)^{\alpha}\ \ ,
			\nonumber\\
 \nabla_F^2{}_{\perp}Z^{\alpha} & = & 
	{}_{\perp}Z^{\beta}
	{}_{\perp}(\nabla_{\beta}\dot{N})^{\alpha} -
	{}_{\perp}R^{\alpha}_{\ \rho\beta\sigma}N^{\rho}N^{\sigma}
	{}_{\perp}Z^{\beta}
	- \frac{1}{N^2}g({}_{\perp}Z,\dot{N})
	{}_{\perp}\dot{N}^{\alpha}\nonumber\\
 & & 	- \frac{1}{N^2}g({}_{\perp}Z,\nabla N^2)
	{}_{\perp}\dot{N}^{\alpha}\ \ ,
\end{eqnarray}
where 
\begin{eqnarray}
 {}_{\perp}(\nabla_{\beta}N)^{\alpha} & \equiv & 
	\nabla_{\beta}N^{\alpha} -
 	\frac{N^{\alpha}N_{\gamma}}{N^2}
	\nabla_{\beta}N^{\gamma}\ \ ,\nonumber\\
 {}_{\perp}(\nabla_{\beta}\dot{N})^{\alpha} & \equiv & 
	\nabla_{\beta}\dot{N}^{\alpha} -
 	\frac{N^{\alpha}N_{\gamma}}{N^2}
	\nabla_{\beta}\dot{N}^{\gamma}\ \ ,\nonumber\\
 {}_{\perp}R^{\alpha}_{\ \rho\beta\sigma} & \equiv & 
	R^{\alpha}_{\ \rho\beta\sigma} -
	\frac{N^{\alpha}N_{\gamma}}{N^2}
	R^{\gamma}_{\ \rho\beta\sigma}\ \ .
\end{eqnarray}


\end{document}